\documentclass[11pt]{article}
\usepackage{newtxtext,newtxmath}
\usepackage[letterpaper,margin=0.75in]{geometry}
\usepackage{graphicx}
\usepackage[font=footnotesize,labelfont=bf]{caption}

\usepackage{authblk}

\makeatletter
\def\@maketitle{%
  \newpage
  \null
  \vskip 2em%
  \begin{center}%
  \let \footer \thanks
    {\LARGE \sffamily \bfseries \@title \par}%
    \vskip 1.5em%
    {\normalsize
      \lineskip .5em%
      \begin{tabular}[t]{c}%
        \@author
      \end{tabular}\par}%
    \vskip 1em%
    {\small \@date}%
  \end{center}%
  \par
  \vskip 1.5em}
\makeatother

\usepackage{titlesec}
\titleformat{\section}
  {\sffamily\large\bfseries}
  {\thesection}
  {1em}
  {}
  
\usepackage{abstract}

\usepackage[numbers,super,sort&compress]{natbib}

\makeatletter
\renewcommand{\fnum@figure}{\textbf{Figure \thefigure}}
\renewcommand{\fnum@table}{\textbf{Table \thetable}}
\makeatother
\usepackage[version=4]{mhchem}
\usepackage[utf8]{inputenc}
\usepackage[T1]{fontenc}
\usepackage[dvipsnames]{xcolor}
\usepackage{dcolumn}
\usepackage{bm}
\usepackage[colorlinks=true, linkcolor=RoyalBlue, citecolor=RoyalBlue]{hyperref}
\usepackage[per-mode=symbol, separate-uncertainty]{siunitx}
\usepackage[capitalise]{cleveref}
\usepackage{enumerate}
\usepackage{lineno}

\usepackage{url}

\usepackage[most]{tcolorbox}

\title{Excitons in van der Waals magnetic materials}
\author[1]{Pratap Chandra Adak\thanks{ padak@ccny.cuny.edu}}
\author[2,3,4]{Florian Dirnberger}
\author[5]{Swagata Acharya}
\author[6]{Akashdeep Kamra}
\author[7,8]{Xiaodong Xu}
\author[1,9]{Vinod M. Menon\thanks{ vmenon@ccny.cuny.edu}}
\affil[1]{Department of Physics, City College of New York, New York, NY 10031, USA.}
\affil[2]{Zentrum für QuantumEngineering (ZQE), Technical University of Munich, Garching, Germany.}
\affil[3]{Department of Physics, TUM School of Natural Sciences, Technical University of Munich, Munich, Germany.}
\affil[4]{Munich Center for Quantum Science and Technology (MCQST), Technical University of Munich, Garching, Germany.}
\affil[5]{Materials, Chemical, and Computational Science Directorate, National Laboratory of the Rockies, Golden, Colorado 80401, USA.}
\affil[6]{Department of Physics and Research Center OPTIMAS, Rheinland-Pfälzische Technische Universität, Kaiserslautern-Landau, 67663 Kaiserslautern, Germany.}
\affil[7]{Department of Physics, University of Washington, Seattle, 98195, WA, USA.}
\affil[8]{Department of Materials Science and Engineering, University of Washington, Seattle, WA 98195, USA.}
\affil [9]{Physics Doctoral Program, Graduate Center of the City University of New York (CUNY), New York, NY 10031, USA.}

\date{}

\begin{document}
\maketitle

\begin{abstract}
\textbf{Two-dimensional magnetic semiconductors provide a unique materials platform in which long-range magnetic order coexists with strongly bound excitons. Because excitonic states and magnetic moments originate from the same electronic orbitals and are coupled through intrinsic exchange interactions, optical excitations in these systems exhibit pronounced sensitivity to magnetic order. Recent experiments have revealed unusually strong magneto-optical responses, as well as direct coupling between excitons and magnons, establishing new routes for controlling light–matter interactions with spin degrees of freedom. This Review surveys key developments in the field, focusing on representative material systems, experimental signatures of exciton–magnetism coupling, and the theoretical frameworks used to describe these phenomena. 
We conclude with perspectives on how this rapidly evolving field could enable next-generation optoelectronic and quantum technologies leveraging the coupled dynamics of light, charge, and spin.}
\end{abstract}

\label{sec:intro}

Quantum materials engineered to interconnect multiple degrees of freedom--charge, spin, lattice, and light--within a single crystalline platform enable novel physics and technological applications. 
Following the isolation of graphene, the past two decades have witnessed two crucial developments in materials science. 
Two-dimensional (2D) semiconductors based on transition metal dichalcogenides (TMDs) have become a platform for robust, room-temperature excitons with large binding energies and valley-specific properties~\cite{xu2014, Manzeli2017, Wang2018}. 
More recently, the discovery of magnetism in monolayers of van der Waals (vdW) materials motivated researchers to further investigate a rich family of 2D magnets~\cite{Mcguire2017, Burch2018, Gibertini2019, Mak2019, Gong2019, huang2020, Zhang2021, Wang2022, Xing2023}. 
At the intersection of two conventional classes of materials--magnets and semiconductors--2D materials have opened a new frontier at which tightly bound excitons coexist and interact with collective spin excitations~\cite{Liu2023, Brennan2024}.

While magnetism is a well-established phenomenon, its interplay with exciton physics remains comparatively unexplored. 
Early insights into exciton–spin coupling came from bulk magnetic insulators and semiconductors such as manganese difluoride, europium chalcogenides, followed by Mn-based dilute magnetic semiconductors (DMS), most notably II–VI alloys~\cite{Sell1967, Wachter1979, Furdyna1988, Dietl2014, Dietl2019}.
In DMS, exchange interaction between the localized and itinerant electrons results in sizable Zeeman splittings and exciton magnetic polarons. 
Magnetic quantum wells and, later, TMDs coupled to magnetic substrates showed analogous exchange-induced exciton shifts~\cite{Zhong2017, Zhao2017, Sierra2021}.
However, these latter proximity-based approaches yielded modest spin splittings and suffered from interfacial disorder. 

\begin{figure}[hbt]
    \centering
    \includegraphics[width=14cm]{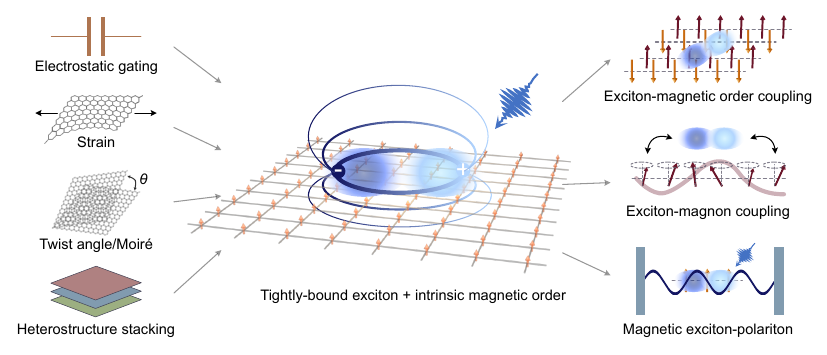}
    \caption{
    \textbf{Exciton-magnetism coupling and tunability in 2D van der Waals magnets.}
    Reduced dimensionality and dielectric screening lead to enhanced excitonic binding energies in layered magnetic materials. External control can be established via electrostatic gating, strain engineering, or by creating moiré superlattices in van der Waals heterostructures. This platform provides a robust framework for investigating enhanced exciton–spin coupling, exciton-magnon interactions, and the emergence of strong light-matter coupling in magnetic exciton-polariton systems.
    \label{fig:intro}}
\end{figure}

The discovery of magnetic 2D semiconductors has renewed and invigorated interest leading to exciting novel insights. 
Because the same transition-metal manifold hosts both the magnetic moments and the electronic transitions forming the exciton, interplay of excitons with magnetism becomes direct and pronounced. 
Rather than acting as a static background, the magnetic order serves as a deterministic switch for excitonic selection rules, spatial confinement, and even optical coherence.
Conversely, excitonic resonances strongly amplify the optical response, enabling sensitive readout of magnetism and its excitations.
Several 2D magnets, such as CrI$_3$~\cite{ Huang2017, Seyler2018, Wu2019, Jin2020}, NiPS$_3$~\cite{Kang2020, Hwangbo2021, Wang2021}, and CrSBr~\cite{Wilson2021, Bae2022, Diederich2023}, already stand out for their extraordinary magneto-optical phenomena demonstrated recently.
Optical properties of some of these materials in their bulk form have already been established~\cite{dillon1966magneto, Piacentini1982}. 
However, their isolation in 2D form has sparked renewed interest.
The 2D architecture offers a dual advantage: built-in exchange fields interact with tightly-bound excitons in reduced dimensionality, while electrostatic gating~\cite{Tabataba2024}, strain~\cite{Cenker2022}, heterostructure and moiré engineering provide external control (Figure~\ref{fig:intro}).

This review examines the novel magneto-optical phenomena that have recently been observed in 2D magnetic semiconductors with a particular focus on excitonic physics. 
We begin by introducing different types of magnetic orders, excitons, and pathways of their interplay in 2D magnetic semiconductors. 
This is followed by a discussion of the magneto-optic phenomena that emerge from this interplay, such as coupling between excitons and spin excitations, and the light-matter interaction underlying exciton-polaritons.
We conclude by outlining emerging challenges and opportunities from both fundamental physics and applications perspectives, demonstrating 2D magnetic semiconductors as a fertile platform for the interplay between light, matter, and magnetism.

\section*{Magnetic and excitonic landscapes}\label{sec:platform}

\paragraph{Spin configurations and their excitations.}
Since the discovery of monolayer magnetism in 2016 and 2017~\cite{Lee2016, Gong2017, Huang2017}, the library of 2D magnetic crystals has expanded rapidly, often exhibiting spin configurations distinct from their bulk counterparts~\cite{Mcguire2017, Burch2018, Gibertini2019, Mak2019}.
Magnetic order in these materials is broadly categorized by two key features: magnetic anisotropy-governed orientation of the spins relative to the crystal plane (described by Ising, XY, or Heisenberg models) and the spin alignment within and between layers, which may be ferromagnetic or various antiferromagnetic types (Box 1).  
Some of these magnets combine robust spin order with strong excitonic resonances, each providing distinct platforms for studying exciton-spin interactions~\cite{Liu2023}. 
Prominent materials include chromium trihalides (\ce{CrX3}, X=Cl,Br,I), transition-metal thiophosphates (\ce{MPS3}, M = Mn, Ni, Fe), and  CrSBr.

\begin{figure}[hbt]
    \centering
    \includegraphics[width=1\linewidth]{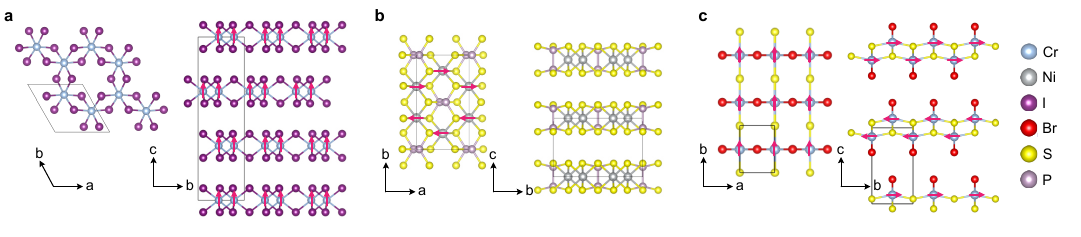}
    \caption{
    \textbf{Crystal structures and magnetic orders of representative 2D magnets.} 
    Top (left) and side (right) views of the crystal structures of \textbf{(a)} \ce{CrI3}, \textbf{(b)} \ce{NiPS3}, and \textbf{(c)} \ce{CrSBr}. 
    The red arrows indicate orientations of magnetic moments of the metal ions.
    In bulk \ce{CrI3}, \ce{Cr^{3+}} spins are ferromagnetically (FM) aligned along the $c$-axis. In the few-layer limit, adjacent layers often exhibit antiferromagnetic (AFM) coupling, while the intralayer coupling remains ferromagnetic, resulting in layer-dependent magnetism. 
    \ce{NiPS3} exhibits an AFM ground state featuring ferromagnetically ordered zigzag chains within the basal plane.
    In \ce{CrSBr}, \ce{Cr^{3+}} spins align along the $b$-axis within each layer, with antiparallel alignment between adjacent layers characteristic of an A-type AFM order. 
    }
    \label{fig:crystal_structures}
\end{figure}

Chromium trihalides consist of layers made of a honeycomb lattice of Cr$^{3+}$ ions, each coordinated by six edge sharing halide octahedra (Fig.~\ref{fig:crystal_structures}a). 
In the bulk, they undergo a structural transition from high-temperature monoclinic to low-temperature rhombohedral stacking.
The magnetic anisotropy varies across the family: a single layer of \ce{CrI3} is a strongly anisotropic ferromagnet with an out-of-plane easy axis, whereas \ce{CrCl3} is an in-plane ferromagnet within each layer.
The interlayer exchange is ferromagnetic in bulk \ce{CrI3} and \ce{CrBr3}, but antiferromagnetic in \ce{CrCl3}~\cite{Wang2019}.
Interestingly, few-layer CrI$_3$ often retains a monoclinic stacking down to low temperature that enforces antiferromagnetic interlayer coupling. 
This leads to a layer-dependent magnetism, i.e., a vanishing net magnetic moment in even-layer samples but a net out-of-plane magnetization in odd-layer samples~\cite{Huang2017}.

In contrast to chromium trihalides, the MPS$_3$ family (M = Ni, Fe, Mn) exhibits a rich  variety of intralayer antiferromagnetic orders~\cite{Joy1992, Sivadas2015}.
The layers consist of a honeycomb lattice of metal cations, each coordinated by edge-sharing \ce{S6} octahedra. 
While MnPS$_3$ is a Néel-AFM with antiparallel nearest-neighbor moments, NiPS$_3$ and FePS$_3$ realize zigzag-AFM order, consisting of ferromagnetic spin chains coupled antiferromagnetically to neighboring chains~\cite{Sivadas2015, Kim2019, Lee2016} (Fig.\ref{fig:crystal_structures}b). 
This quasi-1D spin texture breaks the lattice rotational symmetry, providing the physical foundation for the strongly linearly polarized excitonic emission further discussed below.

The crystal structure of CrSBr is distinct from both the trihalide and thiophosphate families.
It has an orthorhombic structure where each layer is composed of rectangular unit cells (Fig.\ref{fig:crystal_structures}c). 
The Cr$^{3+}$ cations reside within distorted \ce{CrS4Br2} octahedra, which underpins a strong in-plane anisotropy. 
Within each layer, spins align ferromagnetically along the crystallographic $b$-axis.
The layers themselves are coupled antiferromagnetically, resulting in an $A$-type antiferromagnet~\cite{Guo2018, Lee2021}.

Beyond static spin configurations, each magnetic ground state supports its own spectrum of collective spin excitations, known as magnons (quantized spin waves).
In ferromagnets, magnons correspond to coherent precession of spins around their equilibrium orientation.
In antiferromagnets, they emerge as coupled oscillations of the two magnetic sublattices with characteristic optical and acoustic branches.  
The anisotropy of the magnetic order (Ising, XY, or Heisenberg) governs the magnon gap and dispersion, ranging from gapless modes in easy-plane systems to gapped modes in Ising-type magnets.

\paragraph{Types of excitons.}
Having discussed different spin configurations and their collective excitations, we now turn to the optical excitations that couple to them.
Excitons in 2D magnets span a wide range of spatial extension and microscopic origin (Box 2). 
At one end of this spectrum are localized Frenkel excitons, long studied in molecular and solid-state chemistry~\cite{frenkelorig1,frenkelmolecule}.
They largely arise from the localized electronic transitions within the partially filled $d$ states (onsite and $d$-$d$ in character) of the transition-metal ion, described by ligand-field theory and the Tanabe-Sugano diagrams~\cite{SuganoBook, Sugano1954}. 
At the opposite limit are Wannier–Mott excitons, the spatially extended electron–hole pairs that form a hydrogen-like Rydberg series typical of semiconductors~\cite{Wannier1937,Wang2018}. 
Excitons in 2D magnets, however, do not fit neatly into either one of those categories and frequently exhibit mixed character. 

\begin{figure}[hbt]
    \centering
    \includegraphics[width=1\linewidth]{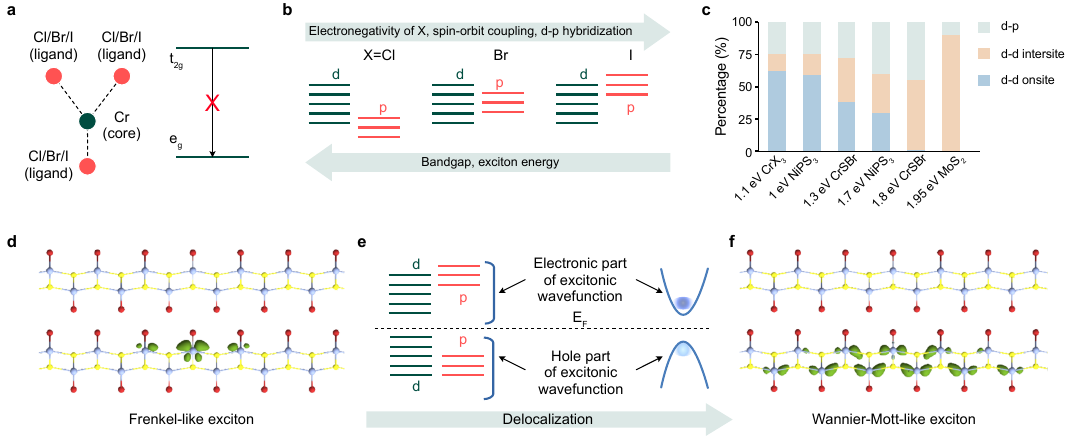}
    \caption{\textbf{Nature of hybrid excitons in 2D magnetic semiconductors.} 
    \textbf{(a)} Schematic of excitonic transitions in chromium trihalides (\ce{CrX3}, \ce{X = Cl, Br, I}), originating from symmetry-forbidden ligand-field $d$--$d$ transitions. 
    \textbf{(b)} Relaxation of selection rules through hybridization between \ce{Cr} $d$-orbitals and halogen $p$-orbitals. This hybridization scales with the halogen’s atomic number and spin-orbit coupling strength (\ce{Cl -> Br -> I}), resulting in enhanced excitonic brightness and Wannier-like delocalization. 
    \textbf{(c)} Comparison of excitonic compositions across various 2D magnets, categorized by $d$--$p$, inter-site $d$--$d$, and on-site $d$--$d$ contributions. Unlike standard Wannier excitons in transition metal dichalcogenides (e.g., \ce{MoS2}), magnetic excitons exhibit significant on-site $d$--$d$ character, though the specific contribution varies across materials. Unlike pure Frenkel excitons, they also possess significant inter-site $d$--$d$ and $d$--$p$ contributions.
    \textbf{(d, f)} Calculated wavefunctions for  \textbf{(d)} 1.37~eV and \textbf{(f)} 1.77~eV excitons of CrSBr, demonstrating a transition from localized Frenkel-like character to delocalized Wannier-like character, respectively. 
    \textbf{(e)} Conceptual illustration of the hybrid exciton continuum, spanning from localized $d$--$d$ transitions toward delocalized band-like states.}
    
    \label{fig:hybrid_excitons}
\end{figure}

Chromium trihalides (CrX$_3$ with X = Cl, Br, I) provide a canonical case, where partially filled $d$ shells and strong electronic correlations favor electronic excitations localized on a single transition-metal ion, while metal–ligand covalency promotes partial delocalization.   
Notably, the main excitonic resonances near $\sim$1~eV remain at comparable energies when going from bulk to the 2D limit, despite changes in dielectric screening.
This can occur because the quasiparticle gap and the exciton binding energy renormalize in opposite directions as screening changes, leading to partial cancellation in the exciton energy~\cite{Zhu2020}.
This relative insensitivity suggests a Frenkel-like assignment for these excitons, arising from parity-forbidden $d$–$d$ transitions between Cr $3d$ $t_{2g}$ and $e_g$ orbitals (Fig.~\ref{fig:hybrid_excitons}a)~\cite{Wu2019,Grzeszczyk2023}. 
However, the hybridization between Cr-3$d$ and halogen-$p$ plays an important role as the primary symmetry breaking mechanism, combined with the lowered symmetry due to the crystal field environment.
This relaxes the Laporte rule and brightens the exciton.
The hybridization increases from Cl to Br to I, as the ligand $p$ states move closer in energy to the Cr $d$ manifold and spin-orbit coupling strength of the ligand atoms enhance (Fig.~\ref{fig:hybrid_excitons}b)~\cite{Molina-Sanchez2020,Acharya2022}. 
Consequently, the lowest excitons become progressively delocalized and intensity becomes brighter along Cl$\rightarrow$Br$\rightarrow$I, while bandgap and exciton binding energy decreases. 
The relative strength of the exciton binding energy to the fundamental gap ($E_B/E_\text{g}$) serves as a useful parameter of the mixed character. In the more ionic CrCl$_3$ (exciton energy $E_X \approx 1.5$~eV, $E_\text{g} \approx 4.5$~eV, $E_\text{b} \approx 3.0$~eV) it is suggestive that the excitonic transition being so strongly bound do not even see the band gap and remains highly atom-local in nature.  
However, in the more covalent CrI$_3$ ($E_X \approx 1.1$~eV, $E_\text{g} \approx 2.4$~eV, $E_\text{b} \approx 1.3$~eV), this ratio drops causing the exciton to acquire a more extended wavefunction and a mixed Frenkel–Wannier character which is sensitive to changes in the band gap and electron and hole properties at the band edges. 

The spatial extent of the exciton wavefunction is fundamentally dictated by its underlying orbital composition: while on-site $d$-$d$ transitions favor more localization, the inclusion of inter-site $d$-$d$ and $d$-$p$ hybridization promotes delocalization across the lattice.
The comparison in Fig.~\ref{fig:hybrid_excitons}c illustrates the spectrum of excitonic character across several 2D magnets, contrasting these hybrid states with spatially extended Wannier–Mott excitons of \ce{MoS2}.
CrSBr offers a particularly clear example of this hybrid behavior in a single material, as it supports multiple excitonic resonances  with varying degrees of localization (Fig.~\ref{fig:hybrid_excitons}d-f).
The lower-energy exciton at 1.36~eV is characterized by a significant on-site $d$-$d$ composition, resulting in a more localized, Frenkel-like state with moderate Wannier character.
In contrast, the higher-energy exciton around 1.77~eV has minimal on-site $d$-$d$ contributions, and exhibit greater delocalization, leading to a predominantly Wannier-like character~\cite{Wilson2021, Datta2025, Shao2025, paulina}. 
This coexistence of different excitons is useful experimentally because it allows one to compare how localization correlates with linewidths, propagation, coupling to phonons, and magneto-optic response within the same host.

In the strongly correlated antiferromagnet NiPS$_3$, the situation is even more complex.
An exciton around $\sim$1.475 eV emerges below the Néel temperature ($\sim$150 K), and is linked to correlated many-body states~\cite{Kang2020}. 
While their precise nature remains under active debate, multiple frameworks have emerged. 
One describes them as Zhang–Rice–type excitons, corresponding to transitions between singlet and triplet many-body states~\cite{Kang2020,Song2024,Klaproth2023}, delocalized over real space and reminiscent of Cooper-pair formation in cuprates~\cite{Zhang1988,Lee2006}. 
Competing models, while corroborating the exciton's singlet nature, propose they are Hund's excitons~\cite{He2024} or excitons formed from largely localized electron–hole correlations on the Ni site with negligible Zhang-Rice component~\cite{jana2025deconstruction}. 
Most key observations in NiPS$_3$ can be interpreted within one or the other framework.

Notably, 2D magnets often host multiple excitons, in contrast to the two primary excitons A and B in TMDs that arise from spin-orbit coupling (SOC) splitting of the valence band. 
However, not all of them acquire sufficient oscillator strength to be prominently observed in the optical spectrum.
Excitons in 2D magnets further differ from those in nonmagnetic TMDs in key ways.
In TMD monolayers, the two Wannier–Mott excitons are formed by interband transitions between delocalized Bloch states at the $K$ and $K'$ valleys~\cite{Wang2018}. 
Their optical selection rules are governed by valley physics, which originates from lattice symmetry and Berry curvature.
Conversely, excitons in 2D magnetic semiconductors are closely tied to the local electronic structure and the underlying magnetic order. 
Their energies and selection rules are shaped by ligand-field multiplet physics, metal–ligand hybridization, and magnetic anisotropy. 
The breaking of TRS and the magnetic point-group symmetry then determines whether circular or linear dichroism is observed.

\paragraph{Pathways for spin-exciton coupling.}
With the diversity of excitonic and magnetic configurations in 2D magnets, we now examine the microscopic pathways through which their mutual coupling can arise. 
Optical excitations primarily probe charge and orbital motion via electric-dipole transitions, while the direct magnetic-dipole coupling of light to spin is very weak. 
Even so, excitons can strongly sense and influence magnetic order whenever magnetism modifies the electronic band structure underlying the excitons~\cite{Wilson2021,Heissenbuttel2025,Semina2025}.

A central mechanism is exchange coupling.
DMS are a well-studied case, where strong $s-d$ and $p-d$ exchange interactions generate sizable Zeeman shifts and splittings of excitonic resonances~\cite{Wachter1979, Furdyna1988, Dietl2014, Dietl2019}.
Such shifts closely follow the magnetization and far exceed the bare spin Zeeman energy.
A defining feature of magnetic van der Waals semiconductors is that both the magnetic order and the lowest-energy optical excitations originate from the same transition-metal–ligand manifold.
The localized $d$-electrons of the transition metal host the magnetic moments, which order via exchange pathways through surrounding ligands. 
Simultaneously, on-site Coulomb interactions and the crystal field split these $d$-orbitals to open a semiconducting gap.
The associated $d-p$ hybridized bands give rise to the dominant excitonic resonances~\cite{Wu2019, Acharya2022, Grzeszczyk2023}. 
This microscopic aspect shared by several vdW magnets contrasts sharply with traditional DMSs, where optical transitions occur within the host $s$- and $p$-derived bands and couple perturbatively to localized magnetic impurities through $sp-d$ exchange. 
In vdW magnets, however, the coupling is intrinsic: optical excitation acts on a correlated electronic background whose spin configuration is part of the excitonic environment itself.

SOC provides another key route.
By mixing spin and orbital degrees of freedom, SOC allows electric-dipole transitions to sense the magnetic state. 
In materials with heavy ligands such as CrI$_3$ and CrBr$_3$, SOC strongly shapes excitonic fine structure and optical selection rules. 
Consequently, changes in magnetic order can switch the polarization of bright excitons, generate pronounced circular dichroism, and induce Zeeman-like splittings.

Magnetostriction and spin–phonon coupling can provide an indirect pathway~\cite{Markina2025}.
Here, magnetic order distorts the lattice or modifies phonon spectra, which in turn shifts exciton energies or brightens otherwise dark states through symmetry lowering.
These indirect effects typically coexist with direct exchange and SOC-driven mechanisms.
Therefore, the overall magneto-excitonic response reflects a combination of spin, orbital, and lattice contributions.

The strength and manifestation of exciton–spin coupling depend sensitively on the exciton type and the magnetic configuration~\cite{Grzeszczyk2023, Brennan2024} and the presence of magnetic disorder~\cite{Ruta2023, Watson2024, Shao2025}.
Factors such as the magnetic easy-axis direction, interlayer coupling, and intralayer spin pattern determine how magnetic order interacts with Frenkel-, Wannier-, Zhang-Rice, or hybrid-type excitons.
The combination of different types of excitons, a wide range of magnetic ground states and spin excitations therefore provides rich grounds for intriguing magneto-optical phenomena.
In the following sections, we discuss how different materials leverage their unique microscopic structures to realize distinct magneto-optic physics.

\section*{Interplay between excitons and static magnetic order}\label{sec:static_MO}

We start by discussing magneto-optical phenomena that arise from the interplay between excitons and magnetic ground states.
In metals, magneto-optic effects arise predominantly from the coupling of light with itinerant carriers near the Fermi level, governed by TRS breaking and SOC.
In 2D magnetic semiconductors, however, strong excitonic resonances dominate the magneto-optic phenomena, resulting in enhanced circular and linear dichroism and many other effects.
The exchange field and SOC can directly modify the orbitals hosting the bound electron–hole pair thereby modifying its energy, oscillator strength, and linewidth.

\paragraph{Exciton-enhanced magnetic circular dichroism.} 

\begin{figure}
    \centering
    \includegraphics[width=1.0\linewidth]{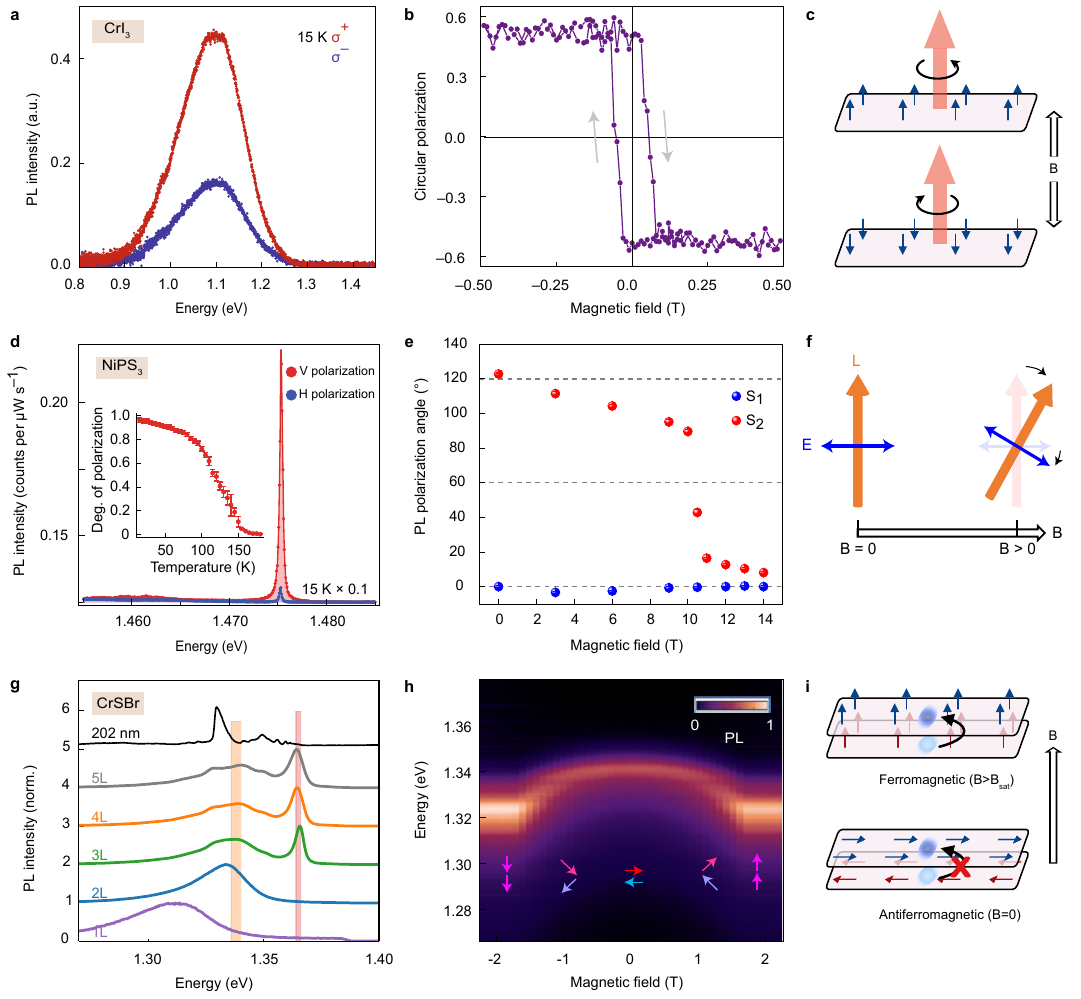}
    \captionof{figure}{
    \textbf{Interplay of excitons with magnetic ground states.}
    \textbf{(a--c)} Magnetic circular dichroism in monolayer \ce{CrI3}. 
    \textbf{(a)} Polarization-resolved PL spectra ($\sigma^+$, red; $\sigma^-$, blue) at 15~K. 
    \textbf{(b)} Circular polarization degree as a function of out-of-plane magnetic field, tracing a ferromagnetic hysteresis loop. 
    \textbf{(c)} Schematic of the circular-polarization selection rules governed by the ferromagnetic spin arrangement. 
    \textbf{(d--f)} Linear dichroism and N\'{e}el vector rotation in \ce{NiPS3}. 
    \textbf{(d)} Linear polarization-resolved PL showing horizontal (H) and vertical (V) components.
    Inset shows the temperature dependence of the degree of polarization. 
    \textbf{(e)} Degree of linear polarization versus in-plane magnetic field for two emission peaks ($S_1, S_2$). Dashed lines indicate the stoichiometric zigzag chain directions. 
    \textbf{(f)} Diagram illustrating the rotation of the linear polarization axis; the N\'{e}el vector ($\mathbf{L}$), electric dipole ($\mathbf{E}$), and magnetic field ($\mathbf{B}$) are shown. 
    \textbf{(g--i)} Magnetism-coupled excitons in \ce{CrSBr}. 
    \textbf{(g)} PL spectra of \ce{CrSBr} as a function of layer number ($N$); shaded regions indicate surface (orange) and bulk (red) excitonic states. 
    \textbf{(h)} Energy redshift from a bilayer sample as an external field along the $c$-axis reduces the magnetic canting angle from the $180^{\circ}$ AFM state. 
    \textbf{(i)} Microscopic mechanism showing spin-dependent electron hopping. Interlayer delocalization is forbidden in the AFM state but becomes allowed as spins align, leading to the observed excitonic redshift.
    Figure adapted from: ref.\cite{Seyler2018}, Springer Nature Ltd (\textbf{a,b}); ref.\cite{Hwangbo2021}, Springer Nature Ltd (\textbf{d}); ref.\cite{Wang2021}, Springer Nature Ltd (\textbf{e,f}); ref.\cite{Shao2025}, Springer Nature Ltd (\textbf{g}); and ref.\cite{Wilson2021}, Springer Nature Ltd (\textbf{h}). 
    }
    \label{fig:exciton_static_magnet}
\end{figure}

Magneto-optic effects--especially the magneto-optic Kerr effect (MOKE) and magnetic circular dichroism (MCD)--have played a key role in the discovery and initial characterization of 2D magnets, where conventional methods established for bulk crystals have proven to be insufficient due to the reduced volume of monolayers. 
For example, MOKE measurements identified long-range ferromagnetic order in monolayer CrI$_3$, along with layer-dependent magnetism: AFM in bilayers, and restored FM in trilayers~\cite{Huang2017}. 
Spontaneous circularly polarized photoluminescence (PL) in monolayer CrI$_3$ directly tracks its magnetic order parameter, with the helicity set by the magnetization direction (Fig.~\ref{fig:exciton_static_magnet}a-c)~\cite{Seyler2018}. 
Reversing the layer magnetization, for example by an external magnetic field, flips the dominant helicity of the circularly polarized emission. 
Notably, observed signals in CrI$_3$ were remarkably strong for atomically thin magnets, far exceeding thickness-scaled estimates from their bulk counterparts.
This enhancement is now understood to be fundamentally excitonic in origin as corroborated by first-principles GW–BSE calculations~\cite{Wu2019, Molina-Sanchez2020, Acharya2022}. 
Unlike its lighter analogues (CrCl$_3$/CrBr$_3$), CrI$_3$ hosts bright excitons with mixed Frenkel–Wannier character with wave functions extending over one to several unit cells. 
This strong excitonic effect was utilized to deterministically flip the magnetization by using circularly polarized light with energy aligned with specific excitonic resonances~\cite{Zhang2022}.
This further solidified the strong excitonic effect determining magneto-optic properties in CrI$_3$.

\paragraph{Magnetic linear dichroism in antiferromagnets.}
In some antiferromagnets such as transition-metal thiophosphates, exciton–spin coupling manifests as a strong magnetic linear dichroism instead of a circular dichroism.
This contrast with bulk ferromagnets, such as CrI$_3$, stems from their distinct lattice and magnetic symmetries. 
A prominent example is NiPS$_3$, a zigzag-type antiferromagnet with an in-plane easy axis and broken rotational symmetry within the layer~\cite{Joy1992, Sivadas2015,  Kim2019}. 
While TRS is locally broken on each sublattice, with zero net magnetization, the system preserves the combined inversion–time-reversal symmetry ($\mathcal{PT}$), which enforces degeneracy between left- and right-circularly polarized transitions.
As a result, circular dichroism is suppressed.
However, the zigzag spin texture breaks the lattice rotational symmetry, lifting the equivalence between orthogonal linear polarizations~\cite{Hwangbo2021, Dirnberger2022}.
Consequently, the excitonic oscillator strength becomes highly anisotropic, producing pronounced magnetic linear dichroism (Fig.~\ref{fig:exciton_static_magnet}d-f). 
Excitons in NiPS$_3$ are strongly polarized orthogonal to the quasi-1D zigzag spin chains (easy axis).
The degree of this polarization closely follows the temperature evolution of the Néel vector, serving as a probe of the symmetry-breaking magnetic order parameter (Fig.~\ref{fig:exciton_static_magnet}d, inset)~\cite{Hwangbo2021, Dirnberger2022}.
Applying a magnetic field rotates the Néel vector and, consequently, the exciton polarization axis, providing a direct proof of its spin-locking (Fig.~\ref{fig:exciton_static_magnet}e-f).

Here we note, CrSBr also shows a pronounced linear dichroism for both emission and reflectance~\cite{Wilson2021}.
In this case, however, the dichroism arises mainly from structural anisotropy, with only a minor magnetic contribution.
For example, in bulk crystals rotating the magnetization direction from $b$-axis to $a$-axis changes linear dichroism only by a few percent. 
Even above the Néel temperature, where magnetic moments are disordered, CrSBr retains strong linear dichroism. 

\paragraph{Magnetic control of exciton energy and confinement.}

In CrSBr, magnetic order directly influences the spatial extent of exciton wave functions.
In this $A$-type antiferromagnet, the intralayer exchange is strong, while interlayer exchange interaction is much weaker\cite{Scheie2022, Lopez-Paz2022}. 
This hierarchy of energy scales produces a magnetically robust system, yet one whose interlayer spin configuration can be readily tuned by modest external magnetic fields \cite{Cham2022}. 
CrSBr's unique low-symmetry structure and orbital configuration, particularly the Cr $d_{yz}$-like valence~\cite{paulina} extending along the $b$-axis and $d_{z^2}$-like conduction edge~\cite{bianchi2023paramagnetic, Watson2024} extending along inter-layer direction, support multiple exciton-spin coupling pathways sensitive to precise nature of the spin alignment~\cite{Brennan2024}. 
Compared to excitons in CrI$_3$ or NiPS$_3$, the excitons in CrSBr possess substantial Wannier character, rendering them spatially extended.
The relative spin alignment between neighboring layers dictates the degree of this delocalization~\cite{Wilson2021,Semina2025,Heissenbuttel2025,Liebich2025,Iakovlev2025}. 
In the AFM state, opposite spin orientations render interlayer electron hopping spin-forbidden, confining the exciton wavefunction within individual layers. 
This magnetic confinement suppresses the layer-number-dependent redshift typically observed in nonmagnetic semiconductors, thereby preserving monolayer-like excitonic properties even in multilayer or bulk crystals~\cite{Shao2025}. 

By contrast, parallel spin alignment in FM state enables interlayer hybridization, increasing exciton delocalization and producing a pronounced redshift of the optical resonance (Fig.~\ref{fig:exciton_static_magnet}h,i). 
The magnitude of this shift depends on the exciton’s degree of Wannier character: the lower-energy exciton near 1.36 eV shifts by roughly 15 meV, while the more delocalized high-energy exciton at $\approx 1.77$ eV exhibits a larger shift approaching 90 meV~\cite{Datta2025,paulina,Komar2024}.
A magnetic field applied along an intermediate or hard axis gradually cants the spins in successive layers toward alignment, allowing the exciton energy to evolve continuously with the field. 
The resulting dependence follows the relation, $\Delta E_X \propto \cos^2(\theta/2)$, where $\theta$ is the canting angle between adjacent layer magnetizations \cite{Wilson2021}. 
The exciton delocalization thus tracks the field-controlled parallel spin component, providing a direct, quantitative link between spin orientation and exciton energy. 

This magnetic confinement is so effective that in few-layer CrSBr, it spectrally differentiates surface excitons from bulk excitons~\cite{Shao2025}. 
The surface excitons, localized to the outermost layers, experience weaker dielectric screening and hence exhibit larger binding energies, appearing at energies below the bulk-like modes (Fig.~\ref{fig:exciton_static_magnet}g). 
In a similar fashion, spin disorder influences excitons in CrSBr.
Between layers, it leads to interlayer hybridization and either a reduction or increase of the exciton energy depending on whether the starting configuration is AFM or FM~\cite{Dirnberger2023}. 
Spin disorder within a single layer, on the other hand, leads to a localization of the exciton wave function that decreases the excitons' oscillator strength~\cite{Ruta2023,Shao2025}. 
CrSBr thus represents a unique platform where the fortuitous combination of $A$-type antiferromagnetism, weak interlayer exchange, and delocalized Wannier-like excitons enables direct and continuous magnetic-field control of exciton properties.

\section*{Dynamical coupling between excitons and magnons}
\label{sec:exciton--magnon}

\paragraph{Optical probing of magnon dynamics.}

Excitons in 2D magnets are not only sensitive to the static spin order but also couple to the collective spin excitations, i.e., magnons.
This dynamical coupling enables a new class of light–spin interaction, where optical responses and spin precession become intertwined.
CrSBr again stands out as the canonical platform, combining strong exciton-spin interaction with low-energy (GHz-frequency) AFM magnon modes that are readily accessible to microwave techniques~\cite{Cham2022}.
Although the energy scales of excitons (eV) and magnons (µeV to meV) differ by several orders of magnitude, their coupling remains remarkably strong, enabling coherent energy exchange between the two quasiparticles.
The mechanism results from the direct energy dependence of excitons on the interlayer magnetic canting angle, $\Delta E_X \propto \cos^2(\theta/2)$.

\begin{figure}
    \centering
    \includegraphics[width=1\linewidth]{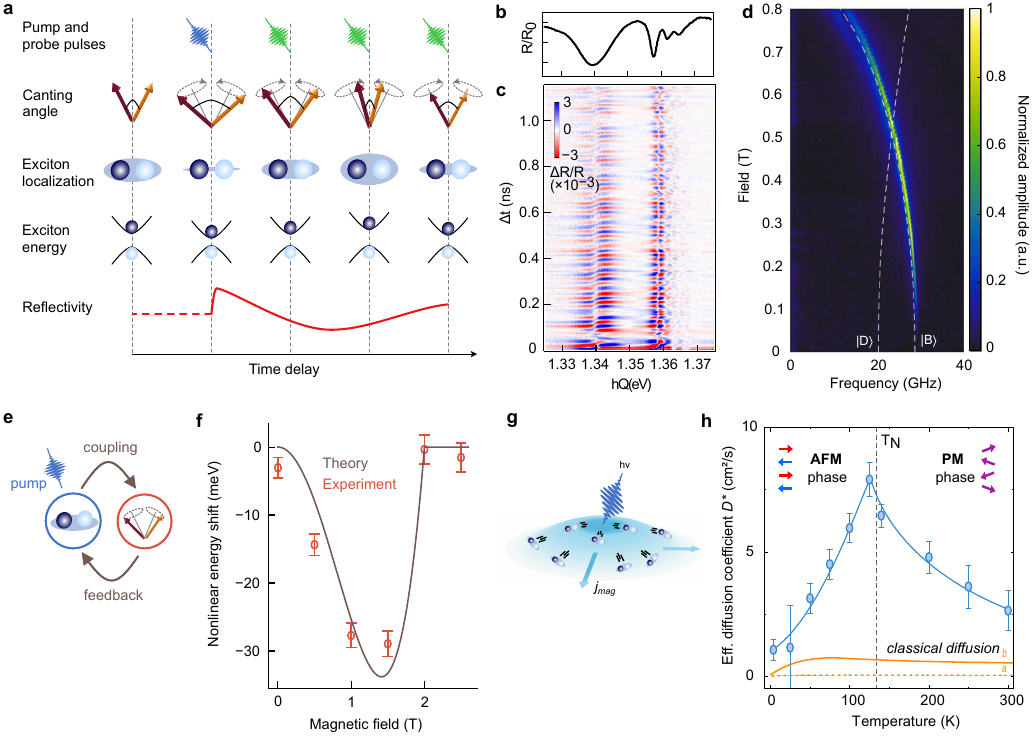}
    \caption{\textbf{Manifestations of exciton--magnon coupling in \ce{CrSBr}.} 
    \textbf{(a)} Schematic of all-optical excitonic detection of magnons. An external magnetic field cants the sublattice spins to an equilibrium angle $\theta_0$. A femtosecond pump pulse generates a population of excitons and coherent magnons; the resulting spin precession modulates the exciton localization and transition energy. This time-resolved energy shift is detected via transient reflectivity ($\Delta R/R$). 
    \textbf{(b)} Reflectance spectrum of \ce{CrSBr} normalized to the \ce{SiO2} substrate. 
    \textbf{(c)} Transient reflectance $\Delta R/R$ as a function of probe energy and delay time, after background subtraction. Periodic oscillations correspond to coherent magnon modes at 24 and 34~GHz. 
    \textbf{(d)} Magnon dispersion with applied magnetic field in \ce{CrSBr} extracted via Fourier transform of the time-domain data. 
    \textbf{(e, f)} Magnon-mediated exciton--exciton interactions. \textbf{(e)} Schematic of the coupling mechanism. \textbf{(f)} Measured nonlinear energy shift of exciton at a fixed fluence as a function of the applied magnetic field; the interaction strength peaks at intermediate magnetic fields and vanishes at zero or saturation fields. 
    \textbf{(g, h)} Magnon-drag effect. \textbf{(g)} Representation of an incoherent magnon flux driving exciton transport. \textbf{(h)} Temperature dependence of the excitonic effective diffusion coefficient, exhibiting a distinct peak at the N\'{e}el temperature ($T_N$), characteristic of magnon--exciton drag effect and contrasting with classical exciton diffusion.
    Figure adapted from: ref.\cite{Bae2022}, Springer Nature Ltd (\textbf{b,c}); ref.\cite{Diederich2023}, Springer Nature Ltd (\textbf{d}); ref.\cite{Datta2025}, Springer Nature Ltd (\textbf{f}); and ref.\cite{Dirnberger2025}, Springer Nature Ltd (\textbf{g,h}). 
    }
    \label{fig:exciton_magnon_coupling}
\end{figure}

Exciton--magnon coupling is linear and larger in the noncollinear (canted) AFM ground state~\cite{Bae2022,Dirnberger2023}, achievable by a modest applied magnetic field.
The weak interlayer exchange in CrSBr, together with comparable magnetic anisotropy, gives rise to two distinct magnon branches in the GHz range that exhibit contrasting optical behavior~\cite{Diederich2023}.
When a magnetic field is applied along the magnetic intermediate ($a$) or hard ($c$) axis, the optical mode, involving out-of-phase precession of spins in adjacent layers, dynamically modulates the canting angle as, $\theta(t)=\theta_0 +\delta \theta (t)$, where $\theta_0$ is the ground-state canting set by an applied magnetic field and $\delta \theta (t)$ represents time-varying oscillations.
This temporal modulation in $\theta$ translates into a periodic shift in exciton energy at the magnon frequency, enabling optical detection of spin-wave dynamics (Fig.~\ref{fig:exciton_magnon_coupling}a-d)~\cite{Bae2022,Dirnberger2023,Sun2024}. 
In time-resolved pump–probe experiments, a femtosecond pump pulse excites coherent magnons, while a delayed probe monitors oscillations in reflectivity at the magnon frequency (Fig.~\ref{fig:exciton_magnon_coupling}a).
The oscillation amplitude peaks at an intermediate magnetic field corresponding to a 90-degree canting, where the linear exciton-magnon coupling is the strongest ($dE_{x}/d\theta_0\propto \sin\theta_0$).
In contrast, the acoustic mode, characterized by in-phase precession of spins on the two sublattices, maintains a constant canting angle and remains optically dark (Fig.~\ref{fig:exciton_magnon_coupling}d). 
Tilting the external magnetic field away from the crystal axis (e.g, introducing a small equilibrium angle $\theta_{ab}$ between the applied field and the $a$-axis) mixes the optical and the acoustic modes.
This mode hybridization leads to an anti-crossing in their dispersion, captured by a beating pattern in the time-resolved optical response~\cite{Diederich2023}, and very large magnon nonlinearities~\cite{Diederich2025}.

While this linear coupling between excitons and magnons in the canted state provides a direct probe of coherent magnetic dynamics, their coupling in the collinear ground state is still effective for other phenomena.
In the collinear ground states, the canting angle averaged over time contains valuable information about the population of magnons~\cite{Dirnberger2023,Dirnberger2025} as well as critical spin fluctuations~\cite{Lopez-Paz2022,Dirnberger2023,Dirnberger2025} close to the ordering temperature.
The coupling in this case is higher order and connects the exciton and magnon densities.
Thus, dynamic detection of exciton energies may provide insights on thermodynamic properties such as magnon temperature and critical exponents, complementary to methods such as qubit spectroscopy~\cite{Casola2018,Du2017} and Brillouin Light Scattering~\cite{Agrawal2013}.

\paragraph{Excitons coupling to spin fluctuations.}
The interaction of excitons with spin fluctuations also raises questions about exciton scattering rates and the evolution of excitonic coherence in magnetic van der Waals materials. 
These issues are particularly relevant in NiPS$_3$, where at very low temperatures the exciton resonance at 1.475\,eV exhibits an exceptionally narrow spectral linewidth of only 0.35\,meV, remarkable for an ensemble-averaged optical response (Fig.~\ref{fig:exciton_static_magnet}d)~\cite{Kang2020}. 
Such a narrow linewidth not only reflects reduced spatial disorder and weak scattering with low-energy phonons, defects, and spin fluctuations, but also points to an unusually long exciton coherence. 
With increasing temperature, the linewidth broadens rapidly, suggesting the onset of efficient scattering with other excitations and disorder-induced decoherence. 
Eventually, this process leads to a depletion of the excitonic resonance from the optical response in the vicinity of the Néel temperature.

Excitonic properties of NiPS$_3$ stand in stark contrast to those of other magnetic van der Waals materials. 
In CrI$_3$, for instance, even at the lowest temperatures exciton linewidths exceeding 100\,meV indicate extremely efficient coupling to phonons~\cite{Seyler2018}. 
In CrSBr, excitonic linewidths are typically on the order of a few meV, which may in part reflect spatial inhomogeneities and disorder, particularly in few-layer samples. 
Nonetheless, in such few-layer samples where additional complications from polaritonic effects can be neglected, careful analyses of temperature-dependent reflectance spectra reveal a clear correlation between exciton linewidth and the AFM-to-PM phase transition~\cite{Shao2025}. 
This behavior provides strong evidence for exciton scattering with both phonons and magnons, although the microscopic dynamics governing these processes remain to be fully understood.

\paragraph{Magnonic control of exciton transport.}
The same coupling that allows optical probing of magnetism, reciprocally, also enables its control, where magnons influence exciton motion and interactions.
The more mobile magnons when thermally driven have been found to drag excitons along by transferring their momentum to the latter in scattering processes (Fig.~\ref{fig:exciton_magnon_coupling}g)~\cite{Dirnberger2025,Iakovlev2025}. 
This magnon-exciton drag peaks around the magnetic critical temperature due to the increased population of the spin excitations at this temperature~(Fig.~\ref{fig:exciton_magnon_coupling}h), in a direct analogy to the corresponding peak observed in studies of magnon-electron drag~\cite{Sugihara1972}. 
Remarkably, the phenomenon does not require a canted magnetic ground state; even a collinear ground state supports the effect~\cite{Sugihara1972,Dirnberger2025,Iakovlev2025}. 
This is because while a noncollinear state supports lower-order 3-particle processes in which an exciton absorbs or emits a magnon, a collinear state admits 4-particle scattering processes between an exciton and a magnon, which are equally, if not more, effective in momentum transfer and causing the drag effect~\cite{Sugihara1972,Dirnberger2025,Iakovlev2025}.

\paragraph{Magnon mediated exciton-exciton interactions.}
As yet another manifestation of the back-action and exciton control, magnetic order has recently been exploited to tune exciton-exciton attraction via an applied magnetic field (Fig.~\ref{fig:exciton_magnon_coupling}e)~\cite{Datta2025}. 
Experimentally, this manifests as a strong excitonic nonlinearity, where the exciton resonance redshifts with increasing pump fluence (exciton density)~\cite{Datta2025}. 
This nonlinearity peaks at intermediate magnetic fields, mirroring the magnetic field dependence of the linear exciton–magnon coupling strength (Fig.~\ref{fig:exciton_magnon_coupling}f). 
The effect is most prominent for Wannier-type excitons, which also exhibit the strongest linear exciton–magnon coupling in time-resolved experiments. 
In essence, a high density of photoexcited excitons adjusts the local spin configuration via exciton–magnon coupling. 
This spin reconfiguration, in turn, shifts the exciton resonance, creating an effective, density-dependent interaction.
This self-consistent feedback between exciton density and spin orientation effectively mediates an exciton–exciton attraction through the magnetic background, tunable via the magnetic ground state and field.
The observation of this effect based on a real magnonic adjustment of the canting angle inspires the feasibility of exciton-exciton interaction via exchange of virtual magnons and quantum spin fluctuations~\cite{Johansen2019}. 
In this scenario, a collinear ground state might already suffice.

\paragraph{Exciton--magnon coupling beyond CrSBr.}
Recent progress on understanding exciton-magnon coupling in CrSBr naturally drives a search for alternate materials with similar physics~\cite{Brennan2024}.
A valuable property of magnons is that they can exhibit high spatiotemporal coherence~\cite{Pirro2021,Sun2024}. 
Time-domain measurements of CrSBr show magnon coherence times exceeding 5 ns, while propagation distances exceed 15 \textmu m, supported by long-range dipolar interactions~\cite{Bae2022, Sun2024}. 
Coupling strength and spectral response can be further tuned using magnetic fields or uniaxial strain~\cite{Diederich2023}.
In contrast, exciton-induced spin dynamics in NiPS$_3$ appear only as short-lived oscillations without persistent coherence~\cite{Belvin2021}.
Crucially, exciton--magnon coupling in CrSBr is underpinned by spatially delocalized excitons in CrSBr, which facilitate spin-resolved interlayer electron and hole hopping. 
$\mathrm{Cr}\mathrm{I}_3$ has been theoretically predicted to also host Wannier excitons besides the more strongly localized, Frenkel-type excitons, which could host similar phenomena~\cite{Wu2019,Olsen2021}. 
However, experiments have so far only found localized Frenkel-like excitons in $\mathrm{Cr}\mathrm{I}_3$~\cite{Seyler2018}, while new excitations continue to be discovered~\cite{He2025}.
$\mathrm{Cr}\mathrm{P}\mathrm{S}_4$ is also expected to host similar excitons and effects with no experimental confirmations yet~\cite{Wu2023, Brennan2024}.

\section*{Strong light-matter coupling and exciton-polaritons}
\label{sec:polaritons}

\begin{figure}[h!t]
    \centering
    \includegraphics[width=1.0\linewidth]{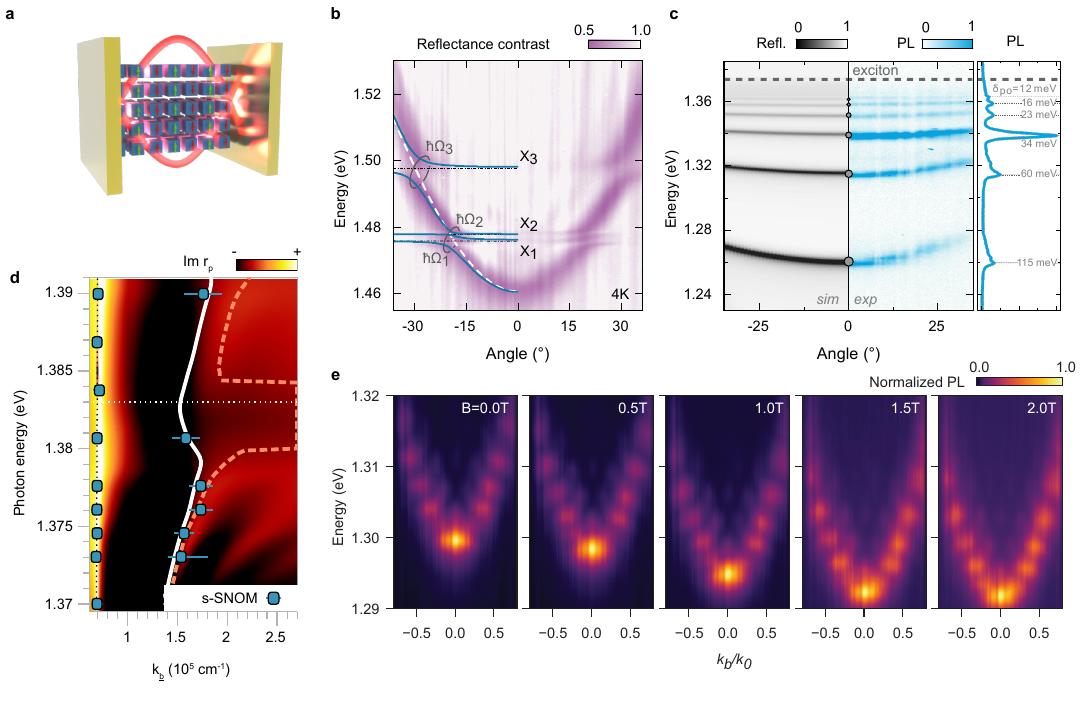}
    \caption{\textbf{Strong light-matter coupling.} 
    \textbf{(a)} Schematic of a layered magnet embedded within an optical microcavity, facilitating the formation of exciton--polaritons. 
    \textbf{(b)} Angle-resolved reflectance contrast map of \ce{NiPS3} at 4~K. Experimental data are overlaid with a coupled-oscillator model (for negative angles) showing the hybridization between the cavity mode and the $X_1, X_2,$ and $X_3$ excitonic resonances. 
    \textbf{(c)} Angle-resolved and angle-integrated photoluminescence (PL) spectra at 1.6~K for a 580-nm-thick CrSBr crystal (right), compared with simulated reflectance (left) showing the polaritonic branches. 
    \textbf{(d)} Momentum-space dispersion of hyperbolic exciton-polaritons (HEPs) in CrSBr and air modes overlaid on the 20~K loss function ($\text{Im}r_p$). The exciton sideband ($X^*$) induces a backbending of the HEP dispersion (white line). 
    \textbf{(e)} Momentum-resolved PL spectra of a micro-cavity-embedded \ce{CrSBr} flake (130~nm thick) for the momentum vector along the $b$-axis. The polaritons inherit magnetic-field-dependent redshift from their excitonic components.
    Figure adapted from: ref.\cite{Dirnberger2022}, Springer Nature Ltd (\textbf{b}); ref.\cite{Diederich2023}, Springer Nature Ltd (\textbf{c}); ref.\cite{Ruta2023}, Springer Nature Ltd (\textbf{d}); and ref.\cite{Adak2025}, Wiley Online Library (\textbf{e}).
    \label{fig:polaritons}}
\end{figure}

As illustrated in \cref{fig:polaritons}a, van der Waals magnetic materials can be readily integrated into optical microcavity resonators. 
This opens a pathway for magnetic systems to enter the regime of strong exciton–photon coupling, in which exciton–polaritons, the hybrid quasiparticles made of both light and matter, are formed.
Experiments investigating magneto-optics in this regime are therefore beginning to address fundamental questions: How do polaritons interact with magnetic order and its excitations? Does their presence alter magneto-optic responses? 

The first experimental realization of strong exciton-photon coupling in a magnetic material was achieved using bulk crystals of NiPS$_3$ embedded in a planar optical microcavity~\cite{Dirnberger2022}. 
The measured polariton dispersion, shown in \cref{fig:polaritons}b, is well described by a coupled oscillator model yielding Rabi splitting energies of only a few millielectronvolts. 
This relatively small coupling strength reflects the modest oscillator strength of bulk excitons in NiPS$_3$, which are believed to be more localized than the Wannier-Mott excitons in monolayer TMDs. 
Consistent with this notion, the study finds polariton-polariton interactions to be weak. 
Besides this result, polariton emission is found to be predominantly linearly polarized and polariton momentum scattering seems to be largely activated by thermal, potentially spin-related processes. 
As a consequence, while strong coupling can be established in NiPS$_3$, its overall impact on magneto-optic properties remains moderate.

A markedly different situation arises in CrSBr, which has emerged as a model system for exploring polaritonic effects in magnetic van der Waals materials~\cite{Dirnberger2023,Wang2023, Adak2025}. 
Here, the large oscillator strength of excitons leads to exceptionally strong light-matter coupling (cf. \cref{fig:polaritons}c). 
Even in the absence of an external cavity, bulk CrSBr supports so-called self-hybridized exciton-polaritons, with coupling strengths manifested via Rabi splitting energies exceeding 200\,meV. 
This establishes a regime in which polaritonic effects dominate the optical response and govern magneto-optic phenomena.

In principle, polaritons inherit characteristics from both their excitonic and photonic constituents. 
In CrSBr, however, magneto-optic effects are found to predominantly originate from the magnetoelectric coupling that governs the properties of the excitons themselves. 
Polaritons nonetheless reshape these responses in a nontrivial way. 
For example, the reflectance modulation induced by coherent magnons is not maximized at the bare exciton resonance, but is instead shifted toward polariton modes with a significant photonic fraction~\cite{Dirnberger2023}.
Moreover, polaritons create magneto-optic responses in spectral regions far beyond the impact of purely excitonic effects. 
An example of this is shown in \cref{fig:polaritons}d, where laterally quantized polariton states induce a magnetic field response far below the exciton energy located near 1.36\,eV.

Besides coherent magnetic excitations, incoherent magnons, spin fluctuations and magnetic disorder have a pronounced influence on the polariton response of CrSBr. 
Disorder-enabled exciton-magnon interactions lead to a reduction of the exciton oscillator strength~\cite{Ruta2023,Shao2025}. 
In the strong coupling regime, this manifests as a quenching of the Rabi gap and, counterintuitively, a substantial upward shift of polariton energies~\cite{Dirnberger2023}. 
This effect is particularly pronounced for polariton states with a large photon fraction, underscoring the delicate balance between magnetic correlations and light-matter hybridization.

The large exciton oscillator strength in CrSBr also has profound consequences for fundamental optical properties, leading to strong anisotropy. 
Excitons strongly renormalize the dielectric response for light polarized along the in-plane crystallographic $b$-axis, giving rise to pronounced interactions between anisotropic optical modes and excitonic resonances~\cite{Li2024}. 
In this regime, hyperbolic polariton modes can emerge, reflecting the highly anisotropic dielectric function. 
Using low-temperature near-field microscopy, Ruta and co-workers directly observed signatures of these hyperbolic modes which are shown in \cref{fig:polaritons}e~\cite{Ruta2023}.

Beyond CrSBr and NiPS$_3$, theoretical work has predicted robust exciton-polaritons in microcavities incorporating other magnetic van der Waals materials, most notably CrI$_3$~\cite{Zhumagulov2023}. 
For monolayer CrI$_3$ embedded in an optical cavity, strong circular dichroism near the exciton resonances leads to helicity-dependent light-matter coupling. 
As a result, cavity photons with opposite circular polarization ($\sigma_+$ and $\sigma_-$) couple with unequal strength to excitonic transitions, giving rise to a zero-field Zeeman splitting of polariton modes that depends on the exciton-cavity detuning and can reach values of up to 20\,meV.
 
\section*{Perspective and outlook}\label{sec:perspectives}

Magnetic materials that also exhibit strong excitonic states in the visible and near-infrared frequency ranges are highly attractive for opto-spintronics. While magnetic materials alone have already found wide applications ranging from data storage to signal processing, the emergence of magnetically ordered systems that are also optically active, such as those highlighted in this review, provides a unique opportunity to merge advances in nanophotonics with long-established magnetic technologies. Moreover, the unique interplay between excitons, magnons, and photons in these materials enables the exploration of novel physical phenomena and the development of device concepts that were previously out of reach. Below, we outline several promising opportunities:

\paragraph{Moiré magneto-excitons.} The relative twist angle between adjacent vdW layers has emerged as a powerful and continuous tuning parameter, capable of dramatically altering electronic, optical, and mechanical properties. The interplay of a material's intrinsic anisotropy and engineered interlayer coupling at different length scales, as dictated by the twist angle, gives rise to a variety of emergent phenomena. Specifically, in the context of magnets, the emerging moiré pattern results in complex magnetic domain structures on demand. Early work on twisted 2D magnets, especially of the chromium trihalide family, showed the possibility to realize nanoscale domains of both FM-favoring and AFM-favoring stacking regions.  Furthermore, these moiré superlattices in twisted magnetic materials offer a perfect environment to generate and stabilize exotic spin textures, skyrmions, which have been considered to be promising information carriers for classical and quantum signal processing~\cite{Psaroudaki2021,Petrovic2025}. 
While most work on moiré magnets in the past was carried out to control magnetic domains, the new class of materials with excitonic states strongly correlated to magnetic order allows one to probe the nanoscale magnetic domains via excitons in the visible and near-infrared spectral range. Here, we note some of the recent discoveries in this context for orthogonal, small-angle, and specific commensurate twist angles. 
The demonstration of multistep magnetization and memory effects in orthogonally twisted CrSBr monolayers~\cite{boix-constant2023} and their imaging using nitrogen vacancy center magnetometry~\cite{healey2024} have been accomplished. 
Recent theoretical works predict the emergence of moiré magnetism and localized excitons in small-angle twisted CrSBr. First-principles calculations have found a critical angle, estimated to be $\theta_c \approx 2.2^\circ$~\cite{Liu2024}, below which ($\theta <\theta_c$) a mixed ground state with coexisting FM and AFM regions is energetically favorable~\cite{Li2025}. 
Furthermore, engineering the interlayer coupling via commensurate twist angles has recently been reported. Specific twist angles seem to create a unique interfacial geometry where the diagonal lattice vectors of the two layers align. First-principles calculations predict that this specific alignment significantly enhances the interlayer orbital overlap and electronic hopping between the layers. An intriguing cross-layer exciton-magnon coupling is also observed around the commensurate angle of 72$^\circ$\cite{sun2025}.
While previous studies on small-angle twisted bilayers of CrI$_3$ found small magnetic domains of FM and AFM configurations coexisting with each other~\cite{Hejazi2020,Song2021,Xu2021,Cheng2023}, a starkly different picture has recently been suggested for CrSBr. Attributed to the significantly stronger intralayer ferromagnetic exchange in CrSBr, a two-sublattice magnetization model assuming a spatially homogeneous configuration adequately reproduces the excitonically measured magnetic states and their switching~\cite{Mondal2025}. The small moiré wavelength of this $\sim 3^\circ$ twisted bilayer further enables a spatial averaging resulting in a twist-controlled effective interlayer exchange of the twisted bilayer~\cite{Mondal2025}.
Topologically protected spin textures, such as skyrmions and merons, which may be stabilized by the induced symmetries and interactions present in these twisted magnetic systems, are yet to be explored, especially their impact on the excitonic properties. Furthermore, one could potentially image these spin textures via excitons.

\paragraph{Optical control of spin textures.} The unique coupling between electronic excitations and spin degrees of freedom in the magnetic vdW materials discussed here naturally raises an important question: Can spin textures be controlled via the polarization state of an excitation laser through the mediation of excitons or exciton-polaritons? Previous demonstrations of optically manipulating magnetic skyrmions have largely relied on ultrashort laser pulses in conventional ferromagnets~\cite{Je2018, Berruto2018, Novakovic-marinkovic2020, Buttner2021, Gerlinger2021, Kern2022}. Magnetic skyrmions themselves are of significant interest for both classical and quantum information processing. The possibility of optically controlling their spatial localization and helicity using the orbital angular momentum (OAM) of light is particularly compelling~\cite{Guan2023}.
In materials such as CrSBr, stronger absorption at exciton and exciton–polariton resonances substantially lowers the threshold for optically exciting magnons, making it feasible to drive spin textures with practical optical pulse energies and simultaneously track their evolution via excitonic transitions.
Indeed, excitonic optical signatures have already been used to monitor magnon propagation in these systems~\cite{Bae2022, Diederich2023}, underscoring the promise of exciton-based approaches for both control and readout of complex spin configurations.

\paragraph{Magneto-photonic devices.} Magneto-optical storage and readout were promising strategies for data storage applications before they were outperformed by other approaches. With strong exciton resonances in vdW materials, it could be attractive to revisit previous concepts for magneto-optic memory. Excitons also bring strongly enhanced optical detection efficiency and new approaches to writing magnetic domains using exciton--magnon coupling. Furthermore, the magnetic properties could be electrically/optically controllable and thus allow hybrid device concepts such as electrical writing of magnetic state and optical read out. Some specific possibilities include the realization of a multilevel memory that relies on multistep magnetization and its optical readout via the excitons, all-optical logic circuits, widely tunable light-emitting diodes that rely on the magneto-chrome effect, and potentially a solid-state laser with magneto-optic functionality. The possibility of modulating the laser at THz frequencies is conceivable through the exciton-magnon coupling in these materials. 

\paragraph{Magneto-exciton polariton condensation.} 
Since the first experimental demonstration of exciton-polariton condensation over 15 years ago in inorganic semiconductors, there have been various other platforms, such as organic molecules, 2D transition metal dichalcogenides, and perovskites where condensation has been demonstrated~\cite{Bellessa2024}. There has been a long standing interest in realizing magneto-polariton condensates that are sensitive to magnetic fields, break time reversal symmetries, and show strong correlation to the underlying magnetic order. 
Recent experiments have observed key signatures of polariton condensation in microcavities containing CrSBr crystals~\cite{Han2025,Zhang2025}. 
A decisive advantage of these materials is that strong excitonic effects persist in bulk crystals, which is highly beneficial for the high-fluence optical excitation typically required to reach the condensation threshold. Efficient heat dissipation in bulk further stabilizes the condensate phase. 
An exciting aspect is the emergence of coherence both at the exciton energy (visible to near infrared) and the magnon energy (GHz to few THz). Such a coherent system will be instrumental in quantum transduction and for realizing practical topological polaritonic systems. 

\paragraph{Quantum Transduction.} A key building block for future quantum networks is a transducer capable of converting quantum information between widely separated frequency ranges~\cite{wehner2018}. Superconducting qubits, which are among the most promising quantum computing platforms, operate in the microwave domain; however, efficient long-distance transfer of quantum information requires conversion of these microwave signals into near-infrared photons that can propagate through low-loss optical fibers. One promising architecture involves coupling microwave photons to magnons within a microwave cavity, using the strong exciton–magnon interaction in magnetic semiconductors as the core transduction mechanism, and finally reading out the converted signal through excitonic or exciton–polariton resonances in the near-infrared. This exciton–magnon pathway offers several advantages over existing approaches, including significantly larger bandwidths (a few MHz) than those achievable in optomechanical transducers and a direct magnon–exciton coupling mechanism that avoids the need for higher-order nonlinear processes typically required in YIG-based systems which in turn significantly reduces the efficiency. Preliminary analyses suggest that, with high-quality optical resonators, low-loss microwave cavities, and the intrinsically narrow linewidths of exciton– and magnon–polaritons, such a hybrid architecture could achieve transduction efficiencies approaching the quantum limit. 

As a final remark, we view excitons in van der Waals magnets as a rapidly expanding frontier.
On the experimental side, a large materials space remains only partially mapped: beyond a few leading systems, systematic control of thickness, stacking, twist, dielectric environment, and electrostatic gating is expected to reveal new opportunities.
On the theoretical side, a central open challenge is to develop a quantitatively predictive description of coupled exciton--spin--lattice dynamics.
Progress along these directions would not only sharpen microscopic understanding, but also provide design rules for magneto-photonic functionality.


\bibliographystyle{naturemag}
\clearpage


\begin{tcolorbox}[enhanced jigsaw,breakable,pad at break*=1mm, colback=yellow!5!white,colframe=yellow!50!black, width=\textwidth,
  colbacktitle=yellow!75!black,
  title= Box I: Magnetic Order and Excitations,every float=\centering]
  \label{box:magnetic_order}

    \textbf{Ferromagnet.} 
    In the ferromagnetic ground state, all spins (magnetic moments) are aligned parallel to each other. An excitation is obtained by flipping one of the spins. This spin-flip, on delocalization over all space, becomes the normal spin-1 bosonic quasiparticle in the ferromagnet -- a magnon~\cite{Holstein1940}. Its classical representation, a spin wave, consists of spins precessing about their equilibrium orientation while maintaining fixed phase difference with their neighbors. It is sometimes called coherent magnon since in a rigorous quantum theory, the classical spin wave dynamics is obtained by considering the underlying bosonic magnon mode to be in a coherent or Glauber state~\cite{Rezende1969,Yuan2022}. Considering interactions other than exchange and Zeeman, eigenexcitations in a ferromagnet come to be made of multiple spin-flips acting together~\cite{Kamra2016,Kamra2020}. In literature, the term ``magnon'' is used loosely and labels a broad range of quasiparticles derived from one or more spin flips, localized or delocalized.  

\begin{center}
    \includegraphics[width=14cm]{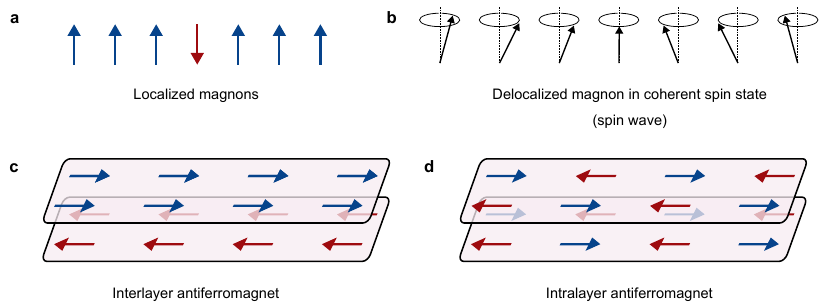} 
    \captionof*{figure}{
    \textbf{Schematic representation of magnons and antiferromagnetic order.} (a) A single spin-flip represents a localized magnon. (b) A delocalized magnon (spin wave) state, where multiple spins precess about their equilibrium orientation with a fixed phase difference between the neighbors. (c) Interlayer antiferromagnetic order, where spins within each layer are aligned parallel. (d) Intralayer antiferromagnet, where spins with antiparallel alignment are present within the layer.
    }
    \label{fig:box_magnetic_order}
\end{center}

\textbf{Antiferromagnet.}
Most van der Waals (vdW) antiferromagnets (AFMs) can be described by a two-sublattice N\'eel ordered ground state~\cite{Burch2018,Wang2022}. Each of the two sublattices effectively forms a ferromagnet with all the spins aligned parallel, while the spins located on opposite sublattices are antiparallel. Several vdW AFMs, such as chromium trihalides~\cite{Cai2019,Huang2018}, CrSBr~\cite{Ziebel2024}, and $\mathrm{Cr}\mathrm{P}\mathrm{S}_4$~\cite{Son2021}, have parallel spins within a layer with interlayer antiferromagnetic coupling. Thus, each physical layer belongs to one sublattice. In contrast, vdW AFMs such as nickel dihalides~\cite{Son2022}, $\mathrm{Ni}\mathrm{P}\mathrm{S}_3$~\cite{Wildes2015}, and $\mathrm{Mn}\mathrm{P}\mathrm{S}_3$~\cite{Ressouche2010}, harbor parallel as well as antiparallel spins within each layer. Thus, each physical layer harbors both the spin sublattices. Since the intralayer (interlayer) coupling is crystalline (vdW), it is strong (weak) and makes these two types of AFMs harbor distinct physics.

Furthermore, due to the strong quantum fluctuations in AFMs, they may harbor complex entangled excitations on top of a many-body ground state~\cite{Savary2017}. Considering even the classical N\'eel ordered ground states~\cite{Rezende2019}, the eigenmode magnons typically consist of several spin flips acting together in an entangled manner~\cite{Kamra2019} and admitting a continuous spin between 0 and 1 determined by the underlying superpositions~\cite{Kamra2017}. Due to the rich nature of the magnonic excitations, their details are system specific.

\end{tcolorbox}

\begin{tcolorbox}[enhanced jigsaw,breakable,pad at break*=1mm, colback=yellow!5!white,colframe=yellow!50!black, width=\textwidth,
  colbacktitle=yellow!75!black,title= Box II: Types of Excitons]

Here, we present rigorous definitions of the three primary exciton classes useful for categorizing optical excitations in van der Waals magnets.

\textbf{Local Frenkel-like multiplet excitons.}
Frenkel excitons~\cite{frenkelorig1, frenkelmolecule} arise from localized electronic transitions, typically onsite $d$–$d$ transitions within a transition-metal ion or its ligand environment. 
Their energies and selection rules are governed by ligand-field theory and Tanabe–Sugano diagrams for a given $d^n$ configuration (e.g., $d^3$ for Cr$^{3+}$, $d^5$ for Mn$^{2+}$, $d^8$ for Ni$^{2+}$ ions)~\cite{SuganoBook, Sugano1954}. 
In centrosymmetric environments, these transitions are typically parity- (Laporte) and spin-forbidden, making them weak or “dark” unless activated by symmetry breaking (e.g., vibronic coupling, distortions, or hybridization). 
The Frenkel description is most accurate when the exciton energy is much smaller than the host bandgap ($E_X/E_\text{g} \ll 1$) (or exciton binding energy is comparable to band gap); a prominent example is ruby (Cr$^{3+}$ in Al$_2$O$_3$). 
While fundamentally intra-ionic, in solids these excitations can couple strongly to lattice vibrations and spin degrees of freedom and may acquire finite dispersion via intersite exchange.

\textbf{Wannier–Mott excitons.}
Conventional semiconductors (e.g., Cu$_2$O, GaAs, GaN, ZnO, TMDs) host Wannier–Mott excitons arising from a conduction-band electron and a valence-band hole bound by the screened Coulomb interaction~\cite{Wannier1937,Wang2018}. 
They are modeled as a hydrogen atom and Rydberg series (with material-dependent effective masses and dielectric screening). 
The exciton radius can extend over many lattice constants, and their energies directly track the band edges ($E_X \approx E_g - E_b$). 
The binding energy $E_b$ typically ranges from $\sim$1–50 meV in bulk crystals to $\sim$0.1-0.5 eV in 2D limits. 
The band-edge dependence distinguishes the Wannier–Mott class from the localized $d$–$d$ transitions discussed above for Frenkel excitons, whose energies are determined by atomic parameters such as crystal-field splitting, Hund’s coupling, on-site Coulomb interactions ($U$), rather than the macroscopic bandgap.

\textbf{Charge-transfer and Zhang–Rice-type excitons.}
Between these limits lie the $p$–$d$ charge-transfer excitons, in which the electron and hole occupy different atoms, sharing partially filled $d$ states of the transition metal and $p$ orbitals of its ligands. 
A prominent subset is Zhang-Rice excitons, in analogy with the original naming of two-particle correlations in cuprates by Zhang and Rice~\cite{Zhang1988, Lee2006}. 

\begin{center}
    \includegraphics[width=16cm]{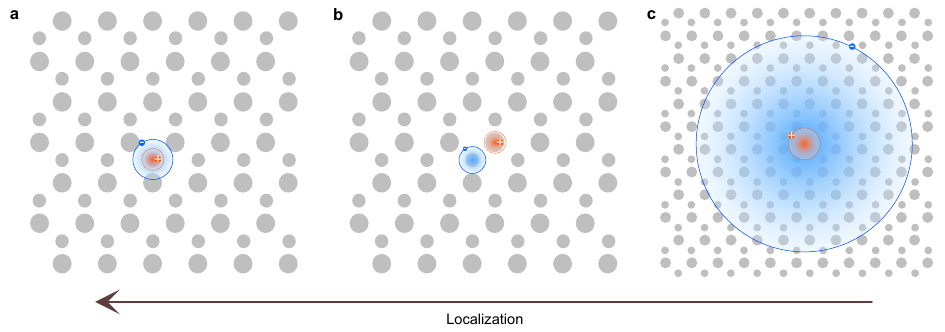} 
    \captionof*{figure}{
    \textbf{Schematic representation of various exciton types.}  Frenkel-like (\textbf{a}), charge-transfer (\textbf{b}), and Wannier-Mott excitons (\textbf{c}).
    }
\label{fig:box_exciton_types}
\end{center}

\textbf{Mixed character and their identification.}
Excitons in correlated 2D magnets often defy simple categorization, mixing local, charge-transfer, and Wannier–Mott character. 
Correctly identifying the dominant character therefore requires theoretical approaches that go beyond simplified band pictures. 
Recent progress has been made using self-consistent Feynman diagram approaches that treat electrons, holes, excitons, and even magnetic excitations (magnons) on equal footing~\cite{acharyaTheoryColorsStrongly2023, questaal_paper, Cunningham2023}. 
However, fully capturing how exciton-spin composites couple to lattice vibrations remains an open challenge.
Experimentally, combining different optical and spin-sensitive techniques is key.
For example, polarization-resolved optical spectroscopy, RIXS, magneto-Raman, and field-dependent photoluminescence help distinguish local from delocalized excitons, reveal ligand-hole character, and correlate excitonic properties with lattice and magnetic order.

\end{tcolorbox}

\clearpage

\section*{Figures}

\clearpage

\section*{Acknowledgments}
This work was authored in part by the National Laboratory of the Rockies for the U.S. Department of Energy (DOE) under Contract No. DE-AC36-08GO28308. 
For S.A., funding was provided by the Computational Chemical Sciences program within the Office of Basic Energy Sciences, U.S. Department of Energy. 
S.A. acknowledges the use computational resources sponsored by the Department of Energy's Office of Energy Efficiency and Renewable Energy and located at the National Laboratory of the Rockies. 
The views expressed in the article do not necessarily represent the views of the DOE or the U.S. Government. 
The U.S. Government retains and the publisher, by accepting the article for publication, acknowledges that the U.S. Government retains a nonexclusive, paid-up, irrevocable, worldwide license to publish or reproduce the published form of this work, or allow others to do so, for U.S. Government purposes.
Financial support by the DFG (German Research Foundation) via the Emmy Noether Program (F.D., Project-ID 534078167), the Munich Center for Quantum Science and Technology (MCQST) under Germany’s Excellence Strategy (F.D. EXC-2111 - 390814868), and Spin+X TRR 173-268565370 (A.K., project A13) is gratefully acknowledged. 
Work at City College of New York was supported by DARPA grant HR0011-25-3-010 (P.C.A) and the Gordon and Betty Moore Foundation grant 12764 (V.M.M).
XX acknowledges the support from the U.S. Department of Energy (DOE), Office of Science, Basic Energy Sciences (BES), under the awards DE-SC0018171 and DE-SC0012509.

\section*{Author contributions}
P.C.A. led the manuscript writing and coordinated discussions, with input from all authors. F.D. contributed to the exciton–polariton section and the discussion of magnetic excitons in NiPS$_3$. S.A. provided the historical and contextual perspective on the classification of magnetic excitons. A.K. contributed to the sections on exciton–magnon coupling in CrSBr and NiPS$_3$, while V.M.M. and X.X. contributed to the Perspective and Outlook section.
All authors commented on the manuscript. V.M.M. initiated and supervised the project.

\section*{Competing interests}
The authors declare no competing interests.

\end{document}